\documentstyle[aps,epsfig]{revtex}
\begin{document}
\draft
\title{Phase separating binary fluids under oscillatory shear}
\author{ Aiguo Xu$^{1}$, G.  Gonnella$^{1}$ and  A. Lamura$^{2}$}
\address{
$^{1}$ Istituto Nazionale per la  Fisica della Materia, Unit\`a di Bari,
{\rm and} Dipartimento di Fisica, Universit\`a di Bari, {\rm and}
TIRES, Center of Innovative Technologies for Signal Detection
and Processing,
 via Amendola 173, 70126 Bari, Italy\\
$^2$ Istituto Applicazioni Calcolo, CNR, Sezione di Bari,\\
Via Amendola 122/I, 70126 Bari, Italy 
}
\maketitle
\begin{abstract}
We apply lattice Boltzmann methods to study the segregation of
binary fluid mixtures under oscillatory shear flow in two dimensions. 
The algorithm allows
to simulate systems whose dynamics is described by the Navier-Stokes and
the convection-diffusion equations. The interplay between several time scales
produces a rich and complex phenomenology. 
We investigate the effects of different
oscillation frequencies and viscosities on the morphology
of the phase separating domains. 
We find that at high frequencies the evolution
is almost isotropic with growth exponents 2/3 and 1/3 in the inertial
(low viscosity) and diffusive (high viscosity) regimes, respectively.
When 
the period of the applied shear flow  becomes  
of the same order of 
the  relaxation time $T_R$ of the
shear velocity profile, 
 anisotropic effects are clearly observable.
In correspondence with non-linear patterns for the velocity profiles, 
we find configurations where lamellar order close to the walls coexists 
with isotropic domains in the middle of the system.
For particular values of frequency and viscosity
it can also happen that 
the convective effects induced by the oscillations
cause an interruption or a slowing of the segregation process,
as found in some experiments. 
Finally, at very low frequencies, the morphology
of domains is characterized by lamellar order everywhere in the system
resembling what happens in the case with steady shear.
\end{abstract}

\section{Introduction}

The process of segregation in fluid mixtures
is greatly affected by the presence of applied flows \cite{Larson,Onuki97}.
The domains of the separating phases  generally grow
with  anisotropic patterns  that  reflect the profile
 of the velocity field.
In the case of  polymer blends subjected to a simple shear flow
string-like structures are observed aligned along the flow
direction \cite{Hashimoto95,Lauger95}. In more complex systems like
diblock copolymer melts  with lamellar order \cite{Bates91}
the question of the orientation of the interfaces is still a debated problem
\cite{Fr95,Morozov02,Corberi02}.
The presence of the flow has also less expected and obvious
consequences. For example, in phase separation of binary fluids,
while in  absence of flow 
the size of  domains  is distributed
around a single  average length-scale \cite{Gunton83}, 
when a shear flow is applied,
two typical lengths are observed for each direction 
\cite{Migler96,Corberi98,Corberi99,Qiu98}.
Another peculiar case is that of  lamellar sheared systems 
where the   symmetry 
of dynamical scaling \cite{Gunton83}, 
generally holding in ordering systems,
is foreseen to be violated \cite{Corberi02}.
Related to the presence of the flow
is also the  behavior of
the stress  response. The non-linear character of this response
reflects the  dynamical evolution of the domain pattern and
is of   fundamental importance
in many technological applications \cite{Tanner88}.

In this paper we study through numerical simulations
the behavior of a symmetric phase separating binary mixture
subjected to an oscillatory  shear flow \cite{Larson}. 
Experiments on this system show that 
in some cases the growth of the domains is interrupted
for frequencies smaller than some critical value 
\cite{Beysens94,Krall93} while in other cases
domains are observed to grow 
on time scales much longer than the period of 
a single oscillation \cite{Matsuzaka97}. Available  simulations 
of this system \cite{Qiu98,Corberi2000,Malevanets99} 
did not take into account 
the  role of hydrodynamics or the existence of a
finite time required to set a linear shear profile 
in a fluid system. 
This time, which is inversely 
proportional to the viscosity in a simple fluid \cite{schlichting},
has a very important role in the case
of oscillatory shear. 
For example, for sufficiently
high frequencies,
as it will be shown later, this time is longer than the oscillation
period 
and the linear profile will be never set in the system.
This will have  relevant consequences 
for the evolution and the morphology of the domains of the separating phases.
Actually, a systematic study of the dependence of the growth properties
 on the 
frequency of oscillations has not been done
in previous simulations  also for the  simplest cases without 
hydrodynamics. 
One has also to observe that, differently from the case
with steady shear, due to the fact that 
the average strain is zero in one period of shearing,
the effects of oscillatory shear on the morphology of domains
are not always easy to understand intuitively.

The  effects discussed above can be properly described only 
considering the full hydrodynamic equations for binary mixtures.
We have used Lattice Boltzmann Method  (LBM) \cite{Benzi92,Chen98,succi}
to  simulate  the   convection-diffusion
and Navier-Stokes equations for a binary fluid.
We have introduced in the lattice Boltzmann scheme appropriate
boundary conditions for a shear flow and we have run our simulations
systematically changing the frequency
of oscillations for a limited set of  values of the  viscosity.
We have considered  the two-dimensional case which 
is also useful 
for the comprehension of the three-dimensional case
and has the advantage of being less demanding from the computational point
of view. 

Lattice Boltzmann simulations are based on discretized Boltzmann 
equations for a set of distribution functions 
related to the fluid densities and velocity.
The densities and momenta are conserved at each step of the 
simulation thus approximating the behavior of the hydrodynamic
equations for the fluid. Lattice Boltzmann methods  
have been found
to be very convenient for simulating quasi-incompressible fluids
on very long time scales, as it is needed in 
phase separation problems \cite{Yeomans,Cates99}. Another advantage 
of the LBM is that, in the implementation of the 
method used in this paper \cite{Osborn95},
a free energy can be  introduced  such that
the fluid relaxes to 
the equilibrium state determined by this free energy.
This allows to know  accurately  the equilibrium properties
of the coexisting phases whose growth dynamics is under study.

Before presenting our results it is useful
to summarize   
the  known behavior of two-dimensional binary mixtures
quenched into the ordered homogeneous phases in absence of flow.
Once domains of the two phases are well established
experimental and numerical data show that the growth is self-similar
with the typical size of domains scaling with time as
$R(t) \sim t^{\alpha}$ \cite{Gunton83}. The growth exponent $\alpha$ 
 depends on the physical mechanism responsible for the phase separation. 
A simple scaling analysis of the Navier-Stokes and of the convection-diffusion 
equations shows that three regimes can be found 
corresponding to the role played by  hydrodynamic degrees of freedom
\cite{Furukawa85,Bray94}.
At high viscosity the domain growth is governed 
by a diffusive mechanism and the growth exponent is 
$\alpha = 1/3$ \cite{Lifshitz61}.
When  hydrodynamics becomes relevant,
the  laws $R(t) \sim t$ or  $R(t) \sim t^{2/3}$
are expected depending on whether viscous forces or inertial 
effects dominate, respectively \cite{Furukawa85,Bray94}. 
In real systems, however, the situation
is more complex. 
The physical mechanism responsible for the viscous growth
is not operating in the two-dimensional case \cite{Siggia79}
and, indeed, this regime
has never been observed in simulations \cite{Yeomans}.

The effects of a steady shear flow
on the growth laws previously discussed have been 
considered in many papers.
In the diffusive case,
analytical calculations based on a self-consistent 
approximation show that the typical size of domains
should grow in the direction normal to the flow as in the case without shear
while the growth exponent in the flow direction
is equal to that in the transverse direction augmented of 
one \cite{Corberi98,Rapapa99}.
This result cannot be easily checked by numerical simulations
due to the presence of finite size effects that become
very soon relevant in the direction of the flow affecting the value
of the exponents \cite{Corberi99}.
The full problem with the Navier-Stokes and the
convection-diffusion equations has been considered in 
Refs.~\cite{Cates99,Rothman91,Padilla97,Lamura00,gonnella,wagner}
but also in this case reliable results for the growth exponents
are not yet available. Actually, the true existence 
of an asymptotic  growth regime in alternative with a stationary 
state with a finite transverse size is a question
still to be clarified \cite{Doi91,Cates99}. 
On the other hand, 
morphological properties are reasonably well understood.
Domains are stretched by the flow and this induces
a coagulation of domains in the flow direction but also ruptures
in the bicontinuous network \cite{Corberi99,Ohta90}. 
As a result, domains assume the typical string-like
character with the already mentioned complication that the size of domains
is  distributed around two typical 
length-scales for each direction.

The lattice Boltzmann scheme used in this paper 
is described in the following Section.
Due to small variations with respect to previous LBM \cite{Orlandini},
details on the method and 
on the implementation are given for convenience of the reader.
The rest of the paper is divided as follows.
In Section III we illustrate our  strategy for the choice of the 
parameters used in the simulations; we also discuss  the relevant  
time scales  for the system considered. In Section IV
 we show  results of simulations where the growth is dominated by
inertial effects while the diffusive case is treated  in Section V.
In Section VI we consider the behavior of the shear stress and we draw 
 our conclusions in Section VII.

\section{The model}

Our simulations are based on the lattice Boltzmann scheme 
developed by Orlandini {\it et al.} \cite{Orlandini} and Swift {\it et al.} 
\cite{Swift}.
In this scheme the equilibrium properties of the system can be controlled
by introducing a free energy which enters properly into the 
lattice Boltzmann model. 

\subsection{The lattice Boltzmann scheme}

The lattice Boltzmann scheme is based on the $D2Q9$ lattice: A
square lattice is used in which each site is connected with nearest and next-to-nearest
neighbors. 
The horizontal and vertical links have length $\Delta x$, the diagonal links
$\sqrt{2} \Delta x$, $\Delta x$ being the space step. 
Two sets of
distribution functions $f_{i}({\bf r}, t)$ and $g_{i}({\bf r}, t)$
are defined on each lattice site ${\bf r}$ at each time $t$.
Each of them is associated with a velocity vector ${\bf e}_i$.
Defined $\Delta t$ as the simulation time step, the quantities
${\bf e}_{i} \Delta t$ are constrained to be  lattice vectors so that 
$|{\bf e}_{i}| = \Delta x/ \Delta t \equiv c$ for 
$i$=1 (East direction), 2 (North), 3 (West), 4 (South)
and $|{\bf e}_{i}| =\sqrt{2} c$ for 
$i$=5 (North East), 6 (North West), 7 (South West), 8 (South East).
Two functions $f_{0}({\bf r}, t)$ and 
$g_{0}({\bf r}, t)$, corresponding to the distribution components that do not
propagate (${\bf e}_{0}={\bf 0}$), are also taken into account.
They evolve during the time step $\Delta t$ according to   
a single relaxation-time Boltzmann equation \cite{bhatnagar,chen1}:
\begin{eqnarray}   
f_{i}({\bf r}+{\bf e}_{i}\Delta t, t+\Delta t)-f_{i}({\bf r}, t)&=&   
-\frac{1}{\tau}[f_{i}({\bf r}, t)-f_{i}^{eq}({\bf r}, t)], \label{dist1}\\   
g_{i}({\bf r}+{\bf e}_{i}\Delta t, t+\Delta t)-g_{i}({\bf r}, t)&=&   
-\frac{1}{\tau_{\varphi}}[g_{i}({\bf r}, t)-g_{i}^{eq}({\bf r},   
t)], \label{dist2} 
\end{eqnarray}   
where $\tau$ and  $\tau_{\varphi}$ are independent   
relaxation parameters, $f_{i}^{eq}({\bf r}, t)$ and 
$g_{i}^{eq}({\bf r}, t)$ are local equilibrium distribution functions.
The distribution functions are related to the total density  {\it n},  
to the fluid momentum $n {\bf v}$ 
and to the density difference $\varphi$ through
\begin{equation}   
n=\sum_{i}f_{i} , \hspace{1.3cm} n{\bf v}=\sum_{i}f_{i}{\bf e}_{i} ,\hspace{1.3cm}   
\varphi=\sum_{i}g_{i} .
\label{phys}   
\end{equation}  

These quantities are locally conserved in any collision process and, 
therefore, we
 require that the local equilibrium distribution functions fulfill the equations 
\begin{eqnarray}
\sum_i (f_i^{eq}-f_i)=0 &\Rightarrow & \sum_i f_i^{eq}=n \nonumber \\
\sum_i (g_i^{eq}-g_i)=0 &\Rightarrow&\sum_i g_i^{eq}=\varphi \label{req1} \\
\sum_i (f_i^{eq}-f_i){\bf e}_i={\bf 0} &\Rightarrow&\sum_i f_i^{eq} {\bf e}_i=
n {\bf v}\nonumber
\end{eqnarray}
Following Refs.~\cite{Orlandini,Swift},
the higher moments of the local 
equilibrium distribution functions are defined so that the 
continuum equations pertinent to a binary fluid mixture can be obtained 
and the equilibrium thermodynamic properties of the system can be controlled.  
We define  
\begin{equation}   
\sum_{i}f_{i}^{eq}e_{i\alpha}e_{i\beta}=c^2 P_{\alpha\beta}+n v_{\alpha} v_{\beta} \;,   
\label{eqn0}
\end{equation}  
\begin{equation}  
 \sum_{i}g_{i}^{eq}e_{i\alpha}=\varphi v_{\alpha} \; ,
\label{eqn} 
\end{equation}
\begin{equation}
\sum_{i}g_{i}^{eq}e_{i\alpha}   
e_{i\beta}=c^2 \Gamma \Delta\mu\delta_{\alpha\beta}+\varphi   
v_{\alpha}v_{\beta} \;.
\label{eqn6}  
\end{equation}   
where $P_{\alpha \beta}$ is the pressure tensor, $\Delta \mu$ is the chemical potential
 difference between the two fluids and
$\Gamma$ is a coefficient related to the mobility of the fluid.
We want to stress that we 
are considering a mixture with the two fluids having the same mechanical
properties and, in particular, the same viscosity. 
The constraint (\ref{eqn}) expresses the fact that  the two 
fluids  have the same velocity. 
The local 
equilibrium distribution functions can be expressed as 
 an expansion at the second order in the 
velocity ${\bf v}$ \cite{Orlandini,Swift}:
\begin{eqnarray}
f_0^{eq}&=& A_0+C_0 v^2 \nonumber\\
f_i^{eq}&=& A_I+B_I v_\alpha e_{i\alpha}+C_I v^2+D_I v_\alpha v_\beta
e_{i\alpha} e_{i\beta}+ G_{I,\alpha\beta}e_{i\alpha} e_{i\beta}
\;\;\;\; i=1,2,3,4 \label{svil1}\\
f_i^{eq}&=& A_{II}+B_{II} v_\alpha e_{i\alpha}+C_{II} v^2
+D_{II} v_\alpha v_\beta
e_{i\alpha} e_{i\beta}+ G_{II,\alpha\beta}e_{i\alpha} e_{i\beta}
\;\;\;\; i=5,6,7,8 \nonumber
\end{eqnarray}
and similarly for the $g_i^{eq}$, $i =0, ..., 8$.
The relations (\ref{req1})-(\ref{eqn0})
can be used to fix the coefficients of these expansions.
A suitable choice of the coefficients in the expansions 
(\ref{svil1}) is
\begin{equation}
A_0=n-20 A_{II}, \hspace{0.5cm} A_I=4 A_{II}, \hspace{0.5cm} 
A_{II}=\frac{P_{\alpha \beta} \delta_{\alpha \beta}}{24}
\label{a's}
\end{equation}
\begin{equation}
B_I=4 B_{II}, \hspace{0.5cm} 
B_{II}=\frac{n}{12 c^2}
\label{b's}
\end{equation}
\begin{equation}
C_0=-\frac{2 n}{3 c^2}, \hspace{0.5cm} C_I=4 C_{II}, \hspace{0.5cm} 
C_{II}=-\frac{n}{24 c^2}
\label{c's}
\end{equation}
\begin{equation}
D_I=4 D_{II}, \hspace{0.5cm} 
D_{II}=\frac{n}{8 c^4}
\label{d's}
\end{equation}
\begin{equation}
\hspace{-2cm} G_{I,\alpha\beta}\!=\!4 G_{II,\alpha\beta}, \hspace{0.5cm}
 G_{II, \alpha \beta}\!=\!\frac{P_{\alpha \beta} 
- \frac{1}{2} P_{\sigma \sigma} \delta_{\alpha \beta}}{8 c^2}
\label{g's}
\end{equation}
The expansion coefficients for the $g_i^{eq}$ can be got from the previous ones
with the formal substitutions $n \rightarrow \varphi$ 
and  $P_{\alpha \beta} \rightarrow \Gamma \Delta \mu \delta_{\alpha \beta}$.
The quantities
 $P_{\alpha \beta}$ and $\Delta \mu$, which appear in the coefficients
of the equilibrium distribution functions, 
can be calculated from a suitable free energy.

\subsection{The equilibrium properties}

The free-energy functional used in the present study is 
\begin{equation}
\begin{cal} F \end{cal}= \int d {\bf r}\left[ \frac{1}{3} n \ln n 
+ \frac{a}{2}\varphi^2
+\frac{b}{4}\varphi^4+\frac{\kappa}{2} (\nabla \varphi)^{2}\right]
\label{fren}
\end{equation}
The term in $n$ gives rise to a positive background pressure and does not 
affect the 
phase behavior; it is required in the lattice Boltzmann 
approach, as it will be seen later. 
The terms in $\varphi$ in the free-energy density $f(n, \varphi)$ correspond 
to the usual 
Ginzburg-Landau free energy typically used in studies of phase separation \cite{Bray94}.
The polynomial terms are related to the bulk
properties of the fluid.  
While the parameter $b$ is always positive,
the sign of $a$ distinguishes between a disordered ($a>0$)
and a segregated mixture ($a<0$) where two pure phases
with $\varphi = \pm \sqrt{-a/b}$ coexist. 
We will consider quenches into the coexistence region 
with $a<0$ and $b=-a$ so that the equilibrium values for the order 
parameter are $\varphi = \pm 1$.
The gradient term is related to the interfacial properties.
The  equilibrium profile between the two coexisting bulk phases
is $\displaystyle \varphi(x)=\tanh \sqrt{\frac{- a}{2 \kappa}}x$ 
giving \cite{rowlinson} a
 surface tension 
\begin{equation}
 \sigma = \frac{2}{3} \sqrt{- 2 a \kappa}
\end{equation}
and  an interfacial width 
\begin{equation}
\xi=2 \sqrt{\frac{ 2 \kappa}{- a}}.
\end{equation}

The thermodynamic properties of the fluid 
follow from the free energy (\ref{fren}).
The chemical potential difference between the two fluids is given by
\begin{equation}
\Delta \mu =  \frac{\delta{\cal F}}{\delta\varphi}=
a \varphi + b \varphi^3 - \kappa \nabla^2 \varphi .
\label{mu}
\end{equation}   
The pressure is a tensor $P_{\alpha\beta}$ since interfaces in the
fluid can exert non-isotropic forces \cite{yang}. The
diagonal part $p_o$ can be calculated from (\ref{fren})
by using thermodynamics relations:
\begin{eqnarray}
p_o&=& 
n \; \frac{\delta{\cal F}}{\delta n}
+\varphi \; \frac{\delta{\cal F}}{\delta\varphi} - f(n, \varphi)\nonumber \\
&=& \frac{1}{3} n + \frac{a}{2}\varphi^2 +\frac{3 b}{4}
\varphi^4 -\kappa \varphi \left( \nabla^2 \varphi \right) -\frac{\kappa}{2}
\left( \nabla \varphi \right)^2
\end{eqnarray}
In deriving the pressure tensor $P_{\alpha\beta}$, one has
to ensure that $P_{\alpha\beta}$ obeys the condition of mechanical equilibrium
$\partial_{\alpha} P_{\alpha\beta}=0$ \cite{evans}. A suitable choice is 
\begin{equation}
P_{\alpha\beta}=p_o \delta_{\alpha\beta}
+ \kappa \partial_{\alpha} \varphi \partial_{\beta} \varphi.
\label{pres}
\end{equation}

The presence of the term depending on $n$ in the free-energy density allows 
to recover the known results of 
the $D2Q9$
lattice Boltzmann model for a single fluid.
Indeed, when
$a=b=\kappa=\varphi=0$, 
the expansion coefficients (\ref{a's})-(\ref{g's})
become those of the $D2Q9$ model \cite{qian}
and
the pressure tensor (\ref{pres}) reduces to the scalar
$p=(c^2/3) n$, 
where we have also included 
the factor $c^2$ appearing in the r.h.s. of Eq.~(\ref{eqn0}).
This is the pressure for an ideal gas with 
speed of sound
$c_s = c/\sqrt{3}$ \cite{qian}. 
Let us observe that the value of the numerical factor in front of
$n \ln n$ in the free energy
depends on the topology of the lattice and the spatial dimensions \cite{qian2}.

It has been shown in Refs.~\cite{Orlandini,Swift}, 
using a Chapman-Enskog expansion \cite{chapman},
that the above 
described lattice Boltzmann scheme simulates at second order in $\Delta t$
the continuity, 
the quasi-incompressible
Navier-Stokes and the convection-diffusion equations
with the kinematic viscosity $\nu$ and the macroscopic mobility $\Theta$
given by
\begin{equation}   
\nu=\Delta t \frac{c^2}{3} (\tau-\frac{1}{2}),
\qquad \Theta=\Gamma \Delta t c^2 (\tau_{\varphi}-\frac{1}{2}).
\label{param}   
\end{equation}   

\subsection{The shear boundary condition}

In order to enforce a shear flow on the system,
we have used the following scheme.
We assume that the shear flow is directed along the horizontal direction.
Boundary walls are placed on the upper and lower rows of sites.
Let us consider the upper wall (similar considerations apply to the lower wall).
After the propagation the distribution
functions $f_0(t)$, $f_1(t)$, $f_5(t)$, $f_2(t)$, $f_6(t)$ and $f_3(t)$ are known on
each site, while $f_7(t)$, $f_4(t)$, $f_8(t)$ are not. 
One uses Eqs.~(\ref{phys}) to determine them as well as $n$.
Requiring that the wall velocities 
$\displaystyle w_{x,t} = \gamma_0 \frac{L-1}{2} \cos (2 \pi f t)$,
$w_{y,t} = 0$
are imposed to the fluid, we can write
\begin{eqnarray}
f_7(t)+f_4(t)+f_8(t)&=&n - \left [f_0(t)+f_1(t)+f_5(t)+f_2(t)+f_6(t)+f_3(t)
\right ] \nonumber \\
f_8(t)-f_7(t)&=&n \;\gamma_0 \;\frac{L-1}{2} \; \cos (2 \pi f t) 
- \left [f_1(t)-f_3(t)+f_5(t)-f_6(t)
\right ] \label{star}\\
f_7(t)+f_4(t)+f_8(t)&=&f_5(t)+f_2(t)+f_6(t)\nonumber
\end{eqnarray} 
where $L$ is the lattice size, $\gamma_0$ is the amplitude of the shear rate 
and $f$ is the frequency of the oscillatory shear.
Consistency of Eqs.~(\ref{star}) gives
\begin{equation}
n = f_0(t)+f_1(t)+f_3(t)+ 2 \left [f_2(t)+f_6(t)+f_5(t)\right ]
\label{cons}
\end{equation}
The system of Eqs.~(\ref{star}) reduces to two equations with three unknown
variables.
To close the system of equations the bounce-back rule 
\cite{lavallee,cornubert} is adopted for  the 
distribution functions
 normal to the boundary. This corresponds to impose that $f_4(t) = f_2(t)$. 
In order to preserve correctly mass conservation we add a further constraint.
Mass will be conserved if the total density 
$n$ on each site is equal to the quantity $\hat{n}$
given by the sum
\begin{eqnarray}
\hat{n}(t,t-\Delta t)\:\:&=&\:\: f_0(t-\Delta t)+f_5(t-\Delta t)+f_2(t-\Delta t)+ 
f_6(t-\Delta t) \nonumber\\
&&+f_1(t)+f_5(t)+f_2(t)+f_6(t)+f_3(t) .
\label{hat}
\end{eqnarray}
where quantities at time $(t-\Delta t)$ refer to the previous time step and have not been
propagated over the lattice.
In order to impose
the constraint that on all the boundary sites $n=\hat n$, we have to introduce
an independent variable in the system of equations. We have chosen $f_0(t)$
since it does not propagate \cite{Lamura00}.
The solutions of the system of Eqs.~(\ref{star}) and $n=\hat n$ are
\begin{eqnarray}
f_0({\bf r},t)&=&\hat{n}- \left [f_1({\bf r},t)+f_3({\bf r},t) \right ]
-2 \left [ f_2({\bf r},t)+
f_5({\bf r},t)+f_6({\bf r},t)\right ] \nonumber \\
f_4({\bf r},t)&=&f_2({\bf r},t) \nonumber \\
f_8({\bf r},t)&=&f_6({\bf r},t)-\frac{1}{2}\left [ f_1({\bf r},t)-f_3({\bf r},t)
\right ]
+ \frac{1}{2} \; n \;\gamma_0 \;\frac{L-1}{2} \cos (2 \pi f t) \label{phi} \\
f_7({\bf r},t)&=&f_5({\bf r},t)+\frac{1}{2}\left [ f_1({\bf r},t)-f_3({\bf r},t)
\right ]
- \frac{1}{2} \; n \;\gamma_0 \;\frac{L-1}{2} \cos (2 \pi f t) \nonumber
\end{eqnarray} 
A similar treatment is required for the $g_i({\bf r},t)$.
With this choice the proper momentum at the
boundary is achieved. At this point the collision step is applied to 
all sites, 
including the boundary ones. 
By this procedure, once the system has been initialized, the application of 
the propagation and collision steps goes on 
preserving mass and momentum conservation and implementing
the correct velocity values on the boundaries, as it has also been
verified numerically \cite{Lamura00}.

Finally, we also require that the two fluids, which, as already stated, 
are assumed to have the same dynamic and static properties, have a neutral wetting with
walls. This can be enforced at each time step by the condition
\begin{equation}
{\bf m} \cdot {\bf \nabla} \varphi = 0
\end{equation}  
where ${\bf m}$ is a unit vector normal to the wall \cite{ludwig,briant}.
This corresponds to fix the gradient of the density $\varphi$ on the walls so that
the angle formed by the  
interfaces between the two fluids with the walls stays at a constant value 
of $\pi/2$ radians.
This completes the description of the model used in the present work.

\section{Parameter selection and relevant time scales}

We have studied  the effects
of the applied flow by  changing the frequency of the oscillations
at fixed values
of the parameters $a,b,\kappa,\tau,\Gamma$. This
  has been systematically done
for the  cases reported in Table~\ref{table1}.
We have kept
the ratio $\kappa/a$ fixed in such a way that the interfacial
width is always of about 3 lattice spacings.
We fixed $\tau_{\varphi}=1 $ with 
$\Gamma$ controlling the value of the macroscopic mobility (\ref{param}).
The amplitude of the shear rate  
is equal in all runs to the value $\gamma_0 = 0.005$.
The size of the lattice, if not otherwise stated,
is $L=256$.

In absence of shear, as observed in \cite{Cates99,catesjfm}, 
the behavior at late times of a viscous phase separating binary mixture
can be described in terms of adimensional temporal and spatial quantities.
Indeed, from the set of macroscopic parameters $n, \sigma, \nu$ 
it is possible to define only one unit of length
 ($L_0 = n \nu^2 / \sigma$) and one of time 
($T_0 = n^2 \nu^3 / \sigma^2$) \footnote{These expressions do not depend on the spatial
dimension.}. 
Then, in a regime of dynamical scaling with 
 the size of domains 
 distributed around a single
typical length, when diffusion is negligible, 
 it is possible to build  up only one spatial and one temporal 
adimensional  variables. These variables have been  used in comparing
results of simulations performed
with  different   parameters and  methods \cite{Cates99,catesjfm}.

The same use of adimensional variables can be done in the case
of very high viscosity when the evolution equations reduce to a  
convection-diffusion equation. $T_D=\xi^3/(\Theta \sigma)$
is the time scale for diffusion \cite{Bray94}.

In the case with oscillatory shear the situation is more complex 
because various temporal scales can be defined and in general one does not
expect dynamical scaling. Due to these reasons we have preferred
to show our results in  terms of original  
quantities. However, 
it remains useful to consider 
the  relevant time scales because they can give informations
 on the physical mechanisms responsible of
phase separation  and on the role of the applied flow.

We have already defined  the quantities $T_0$ and $T_D$ 
which do not depend on the applied flow.
Then there are the time  $T_S = \gamma_0^{-1}$, related to the amplitude 
of the oscillating shear, and the period $T=f^{-1}$ of a single oscillation.
The ratio $T_0/T_S$ can be interpreted as an indicator
of the relative relevance of hydrodynamic and imposed velocities. 
Finally we consider the quantity $T_R = L^2 /(\nu \pi^2) $  
which is the leading contribution to the 
relaxation time for a steady shear profile
in a simple fluid \cite{schlichting}. 
This time can be taken as indicative of the relaxational
velocity processes also in the case of  a phase separating
binary mixture \cite{Lamura00}. 
In Table~\ref{table1} the time scales corresponding to the sets of parameters 
indicated  are also reported. 

The relevance of the ratio $T_R/T$ for the problem considered in this paper
appears clearly from Fig.~1. Here the horizontal velocity $v_x$ profiles
for  the  vertical cross section in the middle of the system
are  reported at a late stage of simulations for the cases 1 and 3 of
Table~\ref{table1} and for two different frequencies.
We checked that these results are independent of the particular
vertical line considered. 
For each case
 the velocity profile is plotted at the quarters of one period.
For the set of parameters 1
with $f=10^{-3}$  the horizontal velocity induced by the shear
is very small in the bulk of the system and comparable with the 
average of the modulus of the vertical
velocity $v_y$.
With the same parameters, when the frequency decreases to $f=10^{-4}$,
the complex non-monotonic behavior of the horizontal velocity
is more evident. $v_x$ is much  larger than hydrodynamic velocities
in a relevant portion of 
the system close to the walls where it also 
assumes opposite directions to those imposed by the walls.
This peculiar pattern of the velocity profiles has consequences
for the behavior of the stress, as it will be seen later.
When the viscosity becomes higher,
like  for the other set of parameters used in Fig.~1, 
the relaxation time
$T_R$ decreases. Then, 
for the same frequencies, the velocity in the bulk
of the system is larger and  almost triangular oscillating
profiles can be  observed at   $f=10^{-4}$ .
We have also checked that each set of four profiles 
of Fig.~1 is typical for the parameters considered 
in the sense that only small variations in the pattern 
of these profiles, probably due to the
evolving interface configuration, can be observed 
during the simulation.
Figure~2 shows the velocity profile 
at the same phase of four  different periods at initial
and late stages of the simulation with set 1 of parameters at $ f = 10^{-3}$.
Small quantitative changes can be observed in the region close to 
the walls while the general shape of  
the profile always remains the same.

Finally we consider the question of the stability 
of the lattice Boltzmann scheme used in this work.
As observed in \cite{catesjfm}, 
this lattice Boltzmann scheme is
intrinsically unstable. At unpredictable times of the simulation
pressure waves grow up indefinitely in very few iterations
making not possible the continuation of the simulation.
As expected, we saw that this problem becomes
more serious 
when $\tau$  tends to the limit 1/2. 
The problem of stability is connected to the one
of guaranteeing as much as possible the
incompressibility of the fluid. Compressibility errors, which go like 
$(v/c_s)^2$ \cite{maier}, 
can be reduced by either increasing $c_s$, which would require to
reduce $\Delta t$, or decreasing the magnitude of 
$a, b, \kappa$ \cite{catesjfm}. 
We have followed a mixed strategy by keeping $\Delta x = 1$ and changing
the values of $\Delta t$ as reported in Table~\ref{table1}. 
In this way we were able to run   simulations long enough
to study the phase separation of  binary mixtures in different 
growth regimes  implicitly 
assuming that the evolution of the  system
is not affected by the possible occurrence of the numerical instability. 
A comment to the results
of Ref.~\cite{catesjfm}  
is that the introduction of walls for the shear 
boundary conditions 
makes worst the stability properties of the LB scheme.

\section{Inertial ordering}

In two-dimensional quiescent systems, as discussed
in the introduction, two growth regimes
with different power law behaviors for the average size of domains $R(t)$
have been clearly identified \footnote{
Growth  regimes with exponent
1/2 have also been reported; their existence at asymptotic times is still 
under debate. 
For a discussion  see  Ref.~\cite{Yeomans}.}.
In this Section we will consider the effects of the oscillatory shear
on the case of phase separation
driven by inertial growth. 
We will mostly refer to the case 1 of Table~\ref{table1}
for which, in absence of flow, the behavior of $R(t)$
 is shown  in Fig.~3.  
The quantity $R(t)$
has been   calculated as 
the first momentum of the structure factor, that is 
\begin{equation}
R(t) = \frac{ \int d k \;\;C(k,t)}{\int d k \;\;k
\;\;C(k,t)}
\end{equation}
where 
 $C(k,t)$ is the spherical average of the structure factor
\begin{equation}
C(\vec k,t)= \langle \varphi(\vec k, t)\varphi(-\vec k, t)\rangle   \qquad 
\end{equation}
and $ \langle \cdot \rangle $ is the average over different histories.
We found a growth exponent 
$\alpha = 0.62 $ 
\footnote {For a more  accurate measure of this exponent we have used
larger lattices with $L = 512$.}; the small discrepancy
from the expected value  $\alpha = 2/3$ 
 typical for the inertial growth can be attributed
to a small violation of dynamical scaling \cite{Wagnerbis}.

The effects of the oscillatory shear on the growth of the domain size
for the case 1 of Table~\ref{table1} can be seen in the panel of Fig.~4 
which summarizes our results for a range 
of
frequencies from $f=10^{-3} $ up to $f = 5 \cdot 10^{-6}$.
We measure the spherical average $R(t)$,
the domain size in the flow direction 
\begin{equation}
R_x(t) = \frac{ \int d\vec k C(\vec k,t)}{\int d\vec k |k_x|
C(\vec k,t)}
\end{equation}
and the analogue $R_y$ for the shear  direction.
The value $f=10^{-3}$ is the highest frequency where
 an anisotropic behavior can be observed.
$R$ and $R_x$ evolve with an  exponent which is 
equal to $2/3 $; 
the change of the slope of
$\log_{10} R$ and $\log_{10} R_x$ at $ \log_{10} t  \sim 4.2$ 
is due to finite size effects which are more relevant in the direction
of the flow. The behavior of
$R_y$ departs from the power law $t^{2/3}$ 
at $\log_{10} t  \sim 3.5$. This anisotropy
is partially due to the presence of the walls
which, even without flow, can inhibit the growth 
of domains in the vertical dimension at the bottom
and at the top of the system.
The morphology of the domains is also influenced
by the  shear velocity which is 
larger compared with the  hydrodynamic velocities
in the region close to the walls - see Fig.~1.
The slightly anisotropic  evolution at $f = 10^{-3}$
can be illustrated  from the two configurations shown in Fig.~5 
with the corresponding structure factors. The circular shape 
of the structure factor at $t= 1650$, whose radius is of the order
of the inverse of $R(t)$, 
reflects  the isotropic  configuration 
of the concentration field at this time. 
At $t = 5900$ the slight prevalence of the peaks
at $k_x = 0 $ corresponds to the presence of a recognizable amount of
domains aligned with the flow close to the walls. 
In the following of this 
simulation the almost isotropic 
character is conserved as it has  been checked looking at the configurations. 

At the frequency $f= 10^{-4}$
 the morphology
of domains is more affected by the applied flow 
due to a bigger region of  the system where
 the horizontal velocity  is significantly larger than the 
typical hydrodynamic velocities measured along the vertical direction - 
see  Fig.~1.
As a consequence lamellae can be observed 
close to the walls
while the growth  keeps a more  isotropic character in the middle 
of the system. Figure~6, at the frequency $f = 2 \cdot 10^{-5}$
 gives  an example of this behavior with
  the evolution of the
system  shown for a whole period.
Larger and more spherical domains can be observed
in the middle of the system
while, close to the walls, thin domains follow the direction of the flow 
and are subjected to a larger number of 
recombination and breakup processes.
In this case and also 
at $f=5 \cdot 10^{-6}$ it is not 
possible to speak of dynamical scaling
since domains are  distributed on different scales.
However, it is worth to observe that
the quantities
 $R$ and $R_y$ follow for a large interval of the
 evolution a power law behavior with exponent 2/3.

When the viscosity becomes larger, as for example in the cases
2,4,5  of Table~\ref{table1}, the evolution of the system without flow
still corresponds to the inertial regime but the resulting
shorter relaxation time $T_R$ makes the presence of the flow in the bulk
more effective with relevant consequences for the kinetics of
phase separation.
In particular we have observed that  the oscillatory
flow can cause the interruption of the segregation process. 
This phenomenon, also reported in
experiments \cite{Beysens94,Krall93}, 
has been found  in our simulations 
at different viscosities and in different growth regimes.
An example of this flow-induced interruption of growth
is shown in Fig.~7.
The growth of $R, R_x, R_y$ becomes very slow at 
$\log_{10} t \sim 4.8$.  In Fig.~7 we also show a set of
 4 configurations in a period 
at this time. We observe that the terminal
regions of domains close to the walls follow with their movement
the oscillation of the flow.  The convection-induced  movements
inhibit
the domain growth  due to other mechanisms (diffusion or inertial)
 and the system 
appears  for a certain interval of time in a sort of 
elastic stationary state. The  size of domains
where this phenomenon is first observed during the phase separation 
is found to 
be of the order of  
the  average deformation $\displaystyle  \frac{2}{L-1}
\int_0^{T/4} dt \int_0^{(L-1)/2} dy \gamma_0 y \cos(2 \pi f t) = 
\gamma_0  \frac{L-1 }{8 \pi} T$
in all cases considered.

\section{Diffusive growth}

In this Section we consider the case where 
diffusion is  the physical mechanism mainly responsible
for phase separation.
We will consider the set 3 of parameters of Table~\ref{table1};
the corresponding behavior of $R(t)$  in quiescent conditions
is shown in Fig.~3 with  the value of the growth
exponent given by $\alpha = 0.35 $.

As in the previous cases also here the growth 
becomes more anisotropic when the frequency $f$ decreases. 
However, due to a higher value of the viscosity,
$T_R$ is smaller and the effects of shear convection are more 
pronounced and observable already for $f=10^{-2}$.
Indeed, at this frequency, and also at
$f=10^{-3}$, as it can be seen in Fig.~8, at late times in the simulation,
 $R_x$  and  $R$ grow faster than $t^{1/3}$ 
with an effective exponent
$\alpha  =  0.39$ at $f= 10^{-2}$ and 
$\alpha  =  0.54$ at $f= 10^{-3}$ for $R(t)$.
This behavior can be understood   by looking at the configurations
of the concentration field. In Fig.~9 it is shown an example
at the time $t=2250$ for $f=10^{-3}$.
Two different phases can be seen to coexist: lamellar ordered domains
aligned with the flow close to the walls and the usual isotropic pattern
of phase separation in the middle of the system.
This coexistence is reflected in the shape of the structure factor
which is circular with two peaks at $k_x =0$ corresponding to the
horizontal  lamellar
domains. Then, as in
diffusive phase
separation with steady shear,  striped domains almost 
aligned with the flow  grow in the flow direction
faster than in the other directions with an exponent
larger than 1/3. In the case of Fig.~9, the effective exponents
for $R_x$ and $R$, which are quantities averaged over the whole system,
will depend on the ratio between  the volumes
of the two coexisting phases. 
 
By  decreasing the frequency, the difference in the
behavior of $R_x$ and $R_y$ becomes more pronounced,
as it can be seen in Fig.~8 at $f= 10^{-4}$ and $f=10^{-5}$.
The four configurations shown in Fig.~10 for the first period of the evolution
at $f=10^{-4}$ exhibit elongated domains in the direction of the flow
similar to those observed in the case of steady shear.
This explains the big difference in the values measured for 
 $R_x$ and $R_y$ in Fig.~8. Of course, also in this case
the morphology of the domains
is strictly related to the behavior of the horizontal velocity
profiles shown in Fig.~1. We see that 
an  almost regular triangular velocity 
profile occurs when 
the ratio between
$T_R$ and $T$ is of  order 1.
Finally, a quantitative evaluation of the growth 
at the late stages of the simulation
at $f=10^{-4}$ can be better done  by 
averaging the behavior of  $R_x$ and $R_y$
over each period. This  gives the exponents $\alpha_x = 0.65$
and  
$\alpha_y = 0.19$. These results  should be compared with the case 
of steady shear where a very slow growth is measured at late times
in  the shear  direction.
 
\section{Stress behavior}

In this Section we consider the behavior of the shear 
stress associated with the deformation of
the domain pattern induced by the flow.
The stress response $\sigma_{xy}$ is calculated 
as the second momentum of the  structure factor:
\begin{equation}
\sigma_{xy} =   \int \frac {d \vec {k}}{(2 \pi)^2} \;\;k_x k_y C(\vec k,t)
\end{equation}

We first discuss a peculiar  behavior that we find for 
the phase of the shear stress.
We show in  Fig.~11
the time evolution  of $\sigma_{xy}$  for the same frequencies and parameters
 of Fig.~4. For convenience   the velocity on the upper wall
is also  plotted in Fig.~11. We observe 
at the frequencies $f = 10^{-3}$ and $f = 10^{-4}$
a phase opposition between the stress and  the velocity field
 imposed on the walls of the system. This unusual phase behavior
can be explained  by looking at the velocity profiles of Fig.~1 where
we see that the velocity not only never relaxes to the triangular
profile but also assumes the ``wrong'' sign in proximity of the walls
before the jump to the values imposed on the boundaries.
Therefore   the stress follows the sign of the real
velocity field in the system and this explains 
the ``strange'' phase behavior of the stress.

The above  analysis is confirmed when we look in Fig.~12
at the case $f = 10^{-4}$
corresponding to  the set 3 of parameters of Table~\ref{table1}.
In this case almost triangular profiles are obtained -
see Fig.~1,  and  
indeed the stress is almost in phase with the velocity field on the walls
\footnote{A complete analysis of the viscoelastic properties of 
phase separating binary mixtures, with the evaluation of the 
elastic and viscous parts of the stress response, is beyond
the purposes of this work.}.

A more general feature of the behavior
of the shear stress can be observed 
in all the cases shown.
We see that  the initial evolution
of the stress  is always characterized by the presence of a peak
which can be eventually followed by  other large
oscillations. This is clear from the
inset of Fig.~11 for  $f= 10^{-3}$
where two overshoots  modulated by small
oscillations due to the velocity field can be observed. 
The phenomenon 
is enhanced at $f= 10^{-4} $ and $f = 2 \cdot 10^{-5}$ 
where the time between successive peaks is of the same order of the period
of the applied flow and is also present at $f = 5 \cdot 10^{-6}$ 
confirming that its origin is independent of the oscillations of the flow.
Indeed, for the latter mentioned frequency,
 we observe overshoots of $\sigma_{xy}$
while the phase of the applied flow has not changed sign.
Similar phenomena are also observed in  Fig.~12.

Overshoots of the shear stress have been reported in experiments
of phase separation with steady shear \cite{Krall92}
 and have also been
found in simulations \cite{Corberi99,Qiu98,Ohta90}. The phenomenon
is interpreted as due to an initial stretching of the domains 
in the direction of the flow to which it corresponds an increase 
of $\sigma_{xy}$. At a certain point the
deformation cannot be sustained 
by the  surface tension and the domains start to break
evolving in less stretched configurations.
This occurs in
 correspondence of a maximum of $\sigma_{xy}$.
Then the system becomes more isotropic 
but, after that a  minimum in $\sigma_{xy}$ is reached, 
elastic energy is again stored 
 due to the deformations and another
overshoot of $\sigma_{xy}$ can be  observed.

The above considerations can be extended also to our
case with further  complications due to the flow  oscillations.
In particular, at frequencies of the order 
of the inverse of time between two overshoots of  $\sigma_{xy}$,
the relaxation or the stretching phenomena discussed above in the case of 
steady shear are greatly influenced by the flow oscillations.
For example, in the case $f = 2 \cdot 10^{-5}$ of Fig.~11,
the relaxation after the first maximum of 
 $\sigma_{xy}$ occurs in correspondence of the change of sign
of the imposed velocity and, therefore, 
one can think that the decreasing
of the stress is mainly due to the reversed sign  of the deformations
than  to breaking processes in the domain pattern.
This results in two very well shaped overshoots than those
occurring in the case  $f = 5 \cdot 10^{-6}$
which resembles what would occur with steady shear.
 
\section{Conclusions}

In this work we  have studied the behavior 
of  phase separating binary mixtures subjected to  oscillatory
shear for different viscosities and frequencies of the applied flow.
The existence
of  different  physical mechanisms operating in phase separation,
the anisotropic effects induced by the flow, and 
  a finite  relaxation time $T_R$
for the triangular velocity profile  contribute to giving  rise 
to a very rich phenomenology.
In particular the role of viscosity is fundamental because 
both the occurrence of
inertial or diffusive growth
 and the time $T_R$ depend on  the viscosity.

In this complex framework  we found that 
the ratio $T_R/T$ can be used as a convenient parameter
for measuring the effects of the applied flow.
At low viscosity and high frequency, for example, 
when the ratio  $T_R/T$ is larger than 1,  
the effects of the shear are limited to a region close 
to the walls of the system, while in the bulk 
the growth is isotropic as in the case without applied flow.
Actually, the most interesting phenomena 
can be observed when the above ratio becomes of order 1.
Different phases, with lamellar order in regions close to the walls
and isotropically oriented domains in the central part of the system,
have been observed to coexist and evolve together in this case.
For particular values of  the frequency and viscosity 
we have also observed the interruption or slowing of the process
of segregation: interfaces are convected successively in opposite directions
with the net effect of inhibiting any other growth mechanism, at least
for a significantly large time interval in simulations.
Finally, 
for values of the ratio $T_R/T$ much less than 1,
 the domains grow with
lamellar morphology everywhere in the system as in the case of steady flow.

The question of the existence of power-law behavior 
for the domain size $R_x,R_y$ has a definite 
answer only in some cases. For sufficiently high frequencies
the power-law behavior of the case without flow, inertial or diffusive
depending on the viscosity, is generally recovered.
At lower frequencies, when different scales for the size of domains are 
observed at the same time in the system,
 it is not possible to speak of dynamical scaling; however,
the quantities $R_x,R_y$ that we measure still give information
about the growth behavior.
We can note that in the limit of very
low frequencies we have not found any signal of a stationary state
with the size of domains reaching or tending to a finite value. 
Actually
we have measured effective growth exponent in the shear direction
less than the expected values in absence of flow  which indicate
a continuation of the 
phase separation also at the late times of our simulations. 
This could be an indication for the case with steady shear
suggesting a growth of domains also at asymptotic times.

We hope that our analysis of the
 segregation process in binary mixtures under oscillatory shear
 will stimulate 
a more systematic experimental investigation of these systems.
A natural continuation of this study will  be its  extension
to the three-dimensional case.

\acknowledgments
A.L. acknowledges INFM for partial support.

\newpage

\begin{table}
\caption{Parameters used in simulations and corresponding time scales.
The shear time scale $T_S$ is equal in all runs to $200$.}
\label{table1}
\end{table}

\begin{figure}
\caption{Horizontal velocity profiles taken at the central vertical
line of the system. The profiles are recorded at the beginning, 
half,  first and third  quarter of the  period considered.}
\label{fig_1}
\end{figure}

\begin{figure}
\caption{The evolution of the horizontal profile
for the case and at the times reported in the inset.
The main frame is a magnification of the 
profiles in the region close to the upper wall.}
\label{fig_2}
\end{figure}

\begin{figure}
\caption{Evolution of the domain size 
in cases without applied flow with periodic boundary conditions. 
The size of the lattice
used in these simulations is $L = 512$.}
\label{fig_3}
\end{figure}

\begin{figure}
\caption{Evolution of the horizontal, vertical, and spherically averaged
 size of domains in double logarithmic scale.
The straight dashed lines have the slope written in the insets.} 
\label{fig_4}
\end{figure}

\begin{figure}
\caption{Configuration of domains with the corresponding structure factor
at two times in a run with parameter Set 1 of parameters at the frequency 
$f = 10^{-3}$. The variables $k_x$ (horizontal axis) and $k_y$ 
(vertical axis) vary in the interval
$[-5/16 \pi, 5/16 \pi]$ and $[- \pi/8, \pi/8]$ 
at the times $t=1650$ and $t=5900$, respectively.}
\label{fig_5}
\end{figure}

\begin{figure}
\caption{Four configurations  and corresponding velocity profiles
in a run with period $T=50000$.}
\label{fig_6}
\end{figure}

\begin{figure}
\caption{Four configurations  and size domain evolution
for a run with parameter Set 4  and $f= 10^{-3}$.
The configurations are taken at the four quarters of a period starting at 
$t = 60000$. The quantities $R_{mx}, R_{my}, R_{m}$
are the horizontal, vertical, spherically averaged domain sizes
averaged over each period.}
\label{fig_7}
\end{figure}

\begin{figure}
\caption{Evolution of the horizontal, vertical, and spherically averaged
domain  size  in double logarithmic scale.
The straight dashed lines have the slope written in the insets.} 
\label{fig_8}
\end{figure}

\begin{figure}
\caption{A configuration with corresponding structure factor
at $t=2250$ in a simulation with parameter Set 3 and $f=10^{-3}$.
In the structure factor plot $k_x$ (horizontal axis)
and $k_y$ (vertical axis) vary in the interval $[-\pi/2 , \pi/2]$.}
\label{fig_9}
\end{figure}

\begin{figure}
\caption{Four configurations in a run  with parameter Set 3 and $f=10^{-4}$.}
\label{fig_10}
\end{figure}

\begin{figure}
\caption{Behavior of the shear stress compared with 
the velocity on the upper wall. In the case with frequency 
$f = 10^{-3}$ the overall behavior is shown in the inset and only the 
initial evolution is plotted in the main frame. The plot of the velocity is
translated along the vertical axis for graphical convenience; units are 
arbitrary for both the quantities.}
\label{fig_11}
\end{figure}

\begin{figure}
\caption{Shear stress and applied flow as in the previous figure, now for
cases with parameter Set 3. In the case with frequency 
$f = 10^{-3}$ the overall behavior is shown in the inset while 
in the main frame the evolution in the time interval $[10000, 20000]$ 
is plotted. The plot of the velocity is
translated along the vertical axis for graphical convenience; units are 
arbitrary for both the quantities.}
\label{fig_12}
\end{figure}

\end{document}